\documentclass[conference]{IEEEtran}
\ifCLASSINFOpdf
\else
\fi

\usepackage{graphicx}
\usepackage{placeins}

\usepackage{textcomp} % for degree symbol

\usepackage{amsmath} % for degree symbol

\begin{document}
\title{High-Sensitivity Iodine Imaging by Combining Spectral CT Technologies }

% \author{\IEEEauthorblockN{Matthew Tivnan, Grace Gang, J. Webster Stayman}
% \IEEEauthorblockA{Department of Biomedical Engineering\\
% Johns Hopkins University}
% \and \IEEEauthorblockN{Wenchao Cao, Nadav Shapria, Peter B. Noël}
% \IEEEauthorblockA{Department of Radiology\\
% Hospital of the University of Pennsylvania } }

\author{Matthew Tivnan\textsuperscript{1}, Grace Gang\textsuperscript{1}, Wenchao Cao\textsuperscript{2}, Nadav Shapira\textsuperscript{2}, Peter B. Noël\textsuperscript{2}, J. Webster Stayman\textsuperscript{1} \\\textsuperscript{1} Johns Hopkins University Department of Biomedical Engineering \\ \textsuperscript{2} Hospital of the University of Pennsylvania Department of Radiology}


%Telephone: (800) 555--1212, Fax: (888) 555--1212}


\maketitle

\begin{abstract}
    Spectral CT offers enhanced material discrimination over single-energy systems and enables quantitative estimation of basis material density images. Water/iodine decomposition in contrast-enhanced CT is one of the most widespread applications of this technology in the clinic. However, low concentrations of iodine can be difficult to estimate accurately, limiting potential clinical applications and/or raising injected contrast agent requirements. We seek high-sensitivity spectral CT system designs which minimize noise in water/iodine density estimates. In this work, we present a model-driven framework for spectral CT system design optimization to maximize material separability. We apply this tool to optimize the sensitivity spectra on a spectral CT test bench using a hybrid design which combines source kVp control and k-edge filtration. Following design optimization, we scanned a water/iodine phantom with the hybrid spectral CT system and performed dose-normalized comparisons to two single-technique designs which use only kVp control or only k-edge filtration. The material decomposition results show that the hybrid system reduces both standard deviation and cross-material noise correlations compared to the designs where the constituent technologies are used individually. 
\end{abstract}


% \fi
\IEEEpeerreviewmaketitle

\section{Introduction}

Spectral CT systems use data acquisition schemes involving varied spectral sensitivities across photon energies. Thus, compared to single-energy CT systems, spectral CT systems can provide more information about the energy-dependent attenuation of the subject being scanned. In particular, spectral CT enables estimation of basis material densities which has many benefits in quantitative clinical imaging. 

Contrast-enhanced imaging of iodine is one of the most widespread clinical applications of spectral CT. Iodine density estimates show contrast-agent concentration, and water density estimates provide virtual non-contrast enhanced images for characterizing patient anatomy \cite{sauter2018dual}. However, the similarity of the attenuation spectra of water and iodine make accurate material decomposition and density estimation challenging. As compared to standard estimation of overall attenuation (as provided by single-energy CT systems), the relative noise is much higher in the individual basis material density estimates provided by spectral CT \cite{faby2015performance}. For some clinical applications, such as imaging pancreatic cancer, very low levels of differential contrast-enhancement can have a meaningful impact on diagnosis \cite{kawamoto2018assessment}. Therefore, development of spectral CT systems which are capable of high-sensitivity water/iodine decomposition is an important goal with direct clinical implications.

One way to improve sensitivity is with advanced data processing. For example, model-based approaches have been widely adopted in single-energy CT for their improved dose-image quality tradeoffs. In this work, we use a direct one-step model-based material decomposition (MBMD) algorithm rather than a two-step reconstruction-decomposition approach, allowing for incorporation of measurement statistics as well as advanced regularization approaches to help reduce noise. 

Another strategy for improving sensitivity, and the focus of this work, is to optimize the spectral CT system design. There are several technologies which can modulate the spectral sensitivity of a CT system: rapid kVp switching \cite{xu2009dual}, multiple x-ray sources \cite{gang2018image}, source filtration (e.g. k-edge filters)  \cite{tivnan2019physical}, dual-layer/multi-layer detectors  \cite{ma2020high}, photon counting detectors \cite{taguchi2013vision}. Each of these spectral modulation technologies have potentially tuneable design parameters such as kVp separation or k-edge filter thickness. There is evidence that combining these methods can improve sensitivity \cite{primak2009improved}. In previous work, we demonstrated that combining spectral modulation technologies into a hybrid system offers greater control over designed spectral sensitivities \cite{tivnan2020combining} \cite{tivnan2019optimized}. Our previous simulation results  showed that joint optimization of these design parameters in a hybrid system can result in higher sensitivity than systems using the constituent spectral modulators individually. 

Recently, we have also proposed a mathematical formula for \emph{material separability index} which models the performance of a spectral CT system and its advantage over single-energy CT. The metric takes into account polyenergetic models of x-ray sources, attenuation, and detector sensitivity, as well as noise and correlations \cite{tivnan2020design}. 

In this work, we apply these theoretical models to the system design optimization of a prototype spectral CT test bench which incorporates both kVp control as well as k-edge filtration. We describe the specific spectral CT model that includes polyenergetic x-ray physics as well as a quantitative metric of water/iodine separability for the modeled system. We describe a spectral CT system design optimization process for a physical prototype including parameters of the design space and optimization methods. Finally, we present the results of a water/iodine imaging study which compares the optimized hybrid kVp control/k-edge filtration design with kVp control individually and k-edge filtration individually.

\section{Methods}

\begin{figure*}[ht]
    \centering
    \includegraphics[trim={00mm 180mm 00mm 0mm},clip,width=0.99\textwidth]{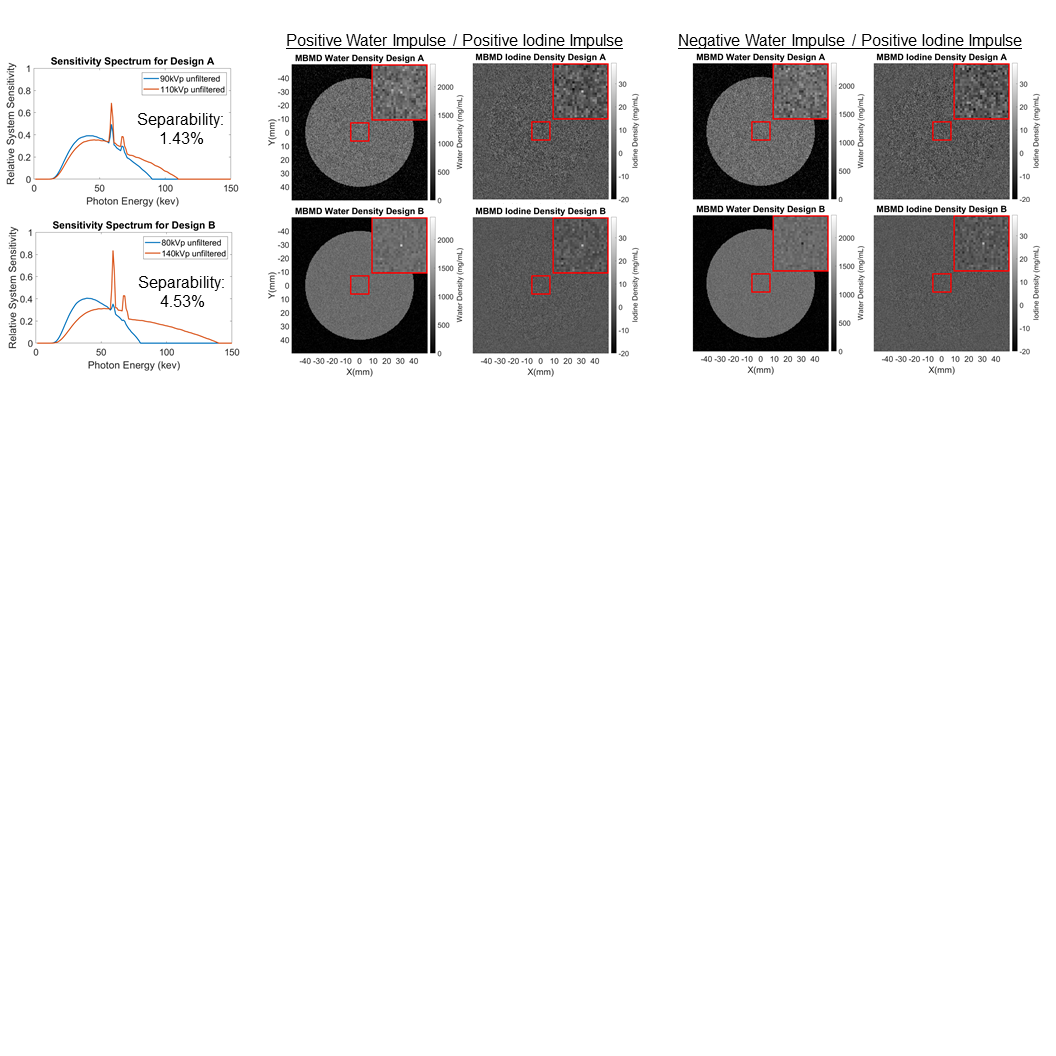}
    \vspace{-3mm}
    \caption{Comparison between two spectral CT system designs. Design B results in a greater separability index than design A. As shown by the basis material densiity estimates, the noise correlation is lower in design B and therefore it is easier to detect a positive impulse of idoine plus a negative impulse of water. } 
    \vspace{-6mm}
    \label{fig:MaterialSeparation}
\end{figure*}

\subsection{Spectral CT Physical Models}

A general forward model for spectral CT is given by:

\vspace{-3mm}
\begin{equation}
    \mathbf{\bar{y}}(\mathbf{x}) = \mathbf{G} \mathbf{S} \exp{ \Big( - \mathbf{Q} \mathbf{A} \mathbf{x} \Big) },
\end{equation}

\vspace{-2mm}

\noindent where $\mathbf{x}$ is a column vector containing basis material densities, $\mathbf{A}$ is a forward projection operator, $\mathbf{Q}$ contains mass attenuation coefficients of basis materials, $\mathbf{S}$ models the system spectral sensitivity, and $\mathbf{G}$ is a gain. This work will focus on modeling flat-panel energy-integrating detectors, for which we can expand $\mathbf{S}$ as follows:
\vspace{-3mm}
\begin{equation}
    \mathbf{\bar{y}}(\mathbf{x}) = \mathbf{G} \mathbf{S_2} \mathbf{S_1} \mathbf{S_0} \exp{ \Big( - \mathbf{Q} \mathbf{A} \mathbf{x} \Big) },
\end{equation}

\vspace{-2mm}

\noindent where $\mathbf{S_0}$ models the spectrum of photons emitted by the source (including filtration), $\mathbf{S_1}$ models the probability of interaction with the scintillator, and $\mathbf{S_2}$ models the generation of secondary quanta in the scintillator, as well as detection and integration by the photodetector. With this spectral CT system model, we can simulate systems with different designs and predict how they will respond to different multi-material objects. For example, we can predict the energy attenuated by an object, $\mathbf{x}$, using the following formula:

\vspace{-3mm}
    \begin{equation}
        \text{Dose}  =  \boldsymbol{\epsilon^T} \mathbf{S_0} \Big(\boldsymbol{1} - \exp{ \big( - \mathbf{Q} \mathbf{A} \mathbf{x} \big)}\Big),
        \label{eq:dose}
    \end{equation}

\vspace{-2mm}

\noindent where $\boldsymbol{\epsilon}$ is a vector containing the energy, in mJ. 

We can also use this model to estimate material density images from spectral CT measurements via MBMD. Assuming a multivariate gaussian noise model with mean, $\mathbf{\bar{y}(x)}$, and covariance, $\boldsymbol{\Sigma_y}$, a maximum-likelihood estimator of basis material densities, $\mathbf{\hat{x}}(\mathbf{y})$, is given by the following formulae:

\vspace{-5mm}
\begin{gather}
        \Phi(\mathbf{x},\mathbf{y}) = \big(\mathbf{y} - \mathbf{\bar{y}}(\mathbf{x})\big)^T \boldsymbol{\Sigma_y^{-1}}\big(\mathbf{y} - \mathbf{\bar{y}}(\mathbf{x})\big) \\
        \mathbf{\hat{x}}(\mathbf{y}) = \underset{\mathbf{x}}{\text{argmin}} \enspace \Phi(\mathbf{x}, \mathbf{y}).
        \label{eq:MBMD}
\end{gather}

\vspace{-2mm}

\noindent Note that the above formula does not include a regularization term. While we focus on unbiased estimators in this work, further noise reduction is possible using the cross-material regularization strategies described in \cite{wang2019generalized} \cite{wang2019local}. To perform the numerical optimization, we use the separable parabolic surrogates algorithm described in \cite{tilley2018model} and a cross material preconditioner defined in \cite{tivnan2020preconditioned} to improve convergence rates. We note that in following sections this MBMD approach will be used for water/iodine decomposition. However, the same estimator can be applied as single-material model-based iterative reconstruction (MBIR) with built-in polyenergetic corrections by using water as the sole basis material.

\subsection{Material Separability}

In this section we present a quantitative metric for material separability originally described in \cite{tivnan2020design}. The metric applies to maximum-likelihood MBMD and is based on the Fisher information matrix which is defined below:

\vspace*{-5mm}
\begin{gather}
    \mathbf{F} = \mathbf{A^T} \mathbf{Q^T} \mathbf{D^T} \mathbf{S^T} \mathbf{G^T} \boldsymbol{\Sigma_y^{-1}} \mathbf{G} \mathbf{S} \mathbf{D} \mathbf{Q} \mathbf{A} \\
    \mathbf{D} = D\{\exp{(-\mathbf{QAx})}\}.
\end{gather}

\vspace{-2mm}

\noindent Note that this expression is object-dependent. That is, different objects, as described by $\mathbf{x}$, will affect the weights in $\mathbf{D}$ and will therefore have an impact on $\mathbf{F}$. This is not surprising since CT image quality is known to be object-dependent. 
%% I don't know that we need this (and we're constrained on space anyway)
%%The Cramer-Rao Lower bound on the covariance of unbiased estimators is given by:
%%
%%\vspace{-5mm}
%%\begin{gather}
%%    \mathbf{\Sigma_x} \leq \mathbf{F}^{-1}.
%%\end{gather}
%%%% WEB - CR is usually defined just for variance.... Need to define \leq for matrices or something here...
%%\vspace{-2mm}
%%%% WEB - ok. careful here. MLE are often asymptotically efficient, but I'm not sure that's guaranteed. There is almost certainly some regularity conditions needed to prove this. Moreover, we know that the general spectral CT forward model is non-convex and that our algorithm doesn't have a perfect convergence proof. I would back off on the following statements about efficiency and/or add the appropriate wiggle words (this audience is more likely to care about these details)
%%\noindent 
%The maximum-likelihood estimator in \eqref{eq:MBMD} is an efficient estimator. That is, it achieves equality for the above expression. 
The detectability index of a signal, $\mathbf{w}$, which is in the same multi-material image space as $\mathbf{x}$, is defined below:

\vspace{-3mm}
\begin{equation}
    d^2(\mathbf{w}) = \mathbf{w^T} \boldsymbol{\Sigma_x^{-1}} \mathbf{w} =  \mathbf{w^T} \mathbf{F} \mathbf{w}.
\end{equation}

\vspace{-2mm}

%%% WEB - Something missing in the following sentence
For two signals, $\mathbf{w_A}$ and $\mathbf{w_B}$ which are normalized such that $d(\mathbf{w_A}) = d(\mathbf{w_B})$, the separability index is defined by the ratio between the detectability of their difference to the detectability of their sum as shown below:

\vspace{-5mm}
\begin{gather}
    s^2(\mathbf{w_A}, \mathbf{w_B}) = \frac{\mathbf{(\mathbf{w_A} - \mathbf{w_B})^T} \mathbf{F} (\mathbf{w_A}-\mathbf{w_B})}{\mathbf{(\mathbf{w_A} + \mathbf{w_B})^T} \mathbf{F} (\mathbf{w_A}+\mathbf{w_B})}.
\end{gather}

\vspace{-2mm}

\noindent In theory this formula can characterize the separability of any two signals, but for the purposes of characterizing water/iodine separability, we may assume that $\mathbf{w_A}$ is a voxel impulse of iodine only at a certain position and $\mathbf{w_B}$ is a voxel impulse of water at the same position. Note that a single-energy CT system would be capable of detecting $(\mathbf{w_A}+\mathbf{w_B})$ but it would be effectively impossible to detect $(\mathbf{w_A}-\mathbf{w_B})$ leading to a material separability index near 0.0\%. In contrast, a spectral CT system with two or more distinct spectral channels should be capable of detecting the differential signal. 
\vspace{-2mm}
\subsection{Spectral CT Prototype System Design Optimization}

Our goal in this section is to apply the theoretical models above to maximize water/iodine separability in a prototype spectral CT test bench using a combination of kVp control and k-edge filtration. The spectral sensitivity design consists of two channels. Each channel was parameterized by three quantities: kVp setting, filtration material, and exposure. We explored six possible kVp settings: 70, 80, 90, 100, 110, and 120~kVp. Eight possible filter materials and thicknesses were 250~$\mu$m praseodymium, 250~$\mu$m erbium, 127~$\mu$m lutetium, 250~$\mu$m hafnium, 100~$\mu$m tungsten, 100~$\mu$m gold, 250~$\mu$m lead, and no filter. For each spectrum there are 48 unique combinations of filter material and kVp settings. Therefore, there are 1128 unique spectral profiles. Exposure settings were optimized in a nested optimization for each possible shape.

\begin{figure}[ht]
    \centering
    \includegraphics[trim={50mm 173mm 50mm 18mm},clip,width=0.49\textwidth]{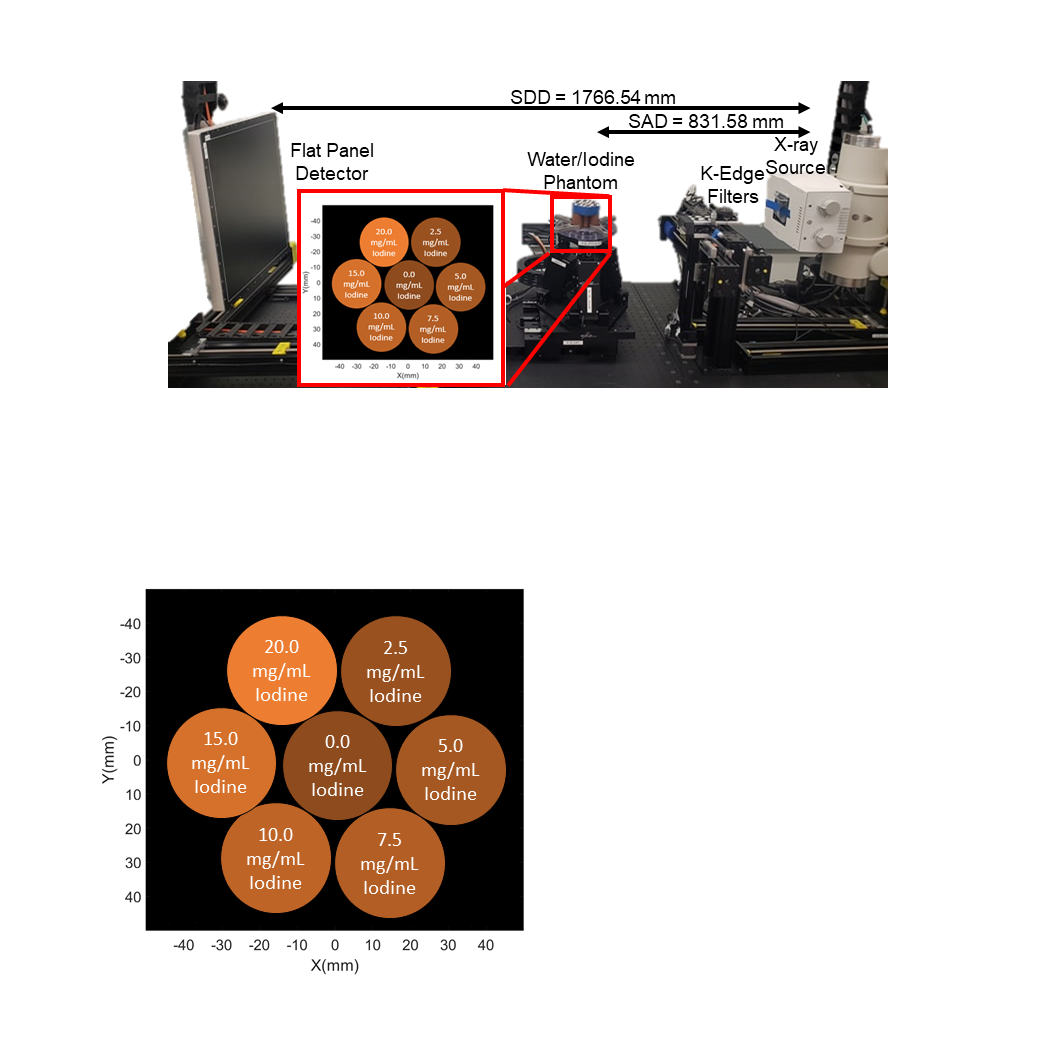}
    \vspace{-8mm}
    \caption{Hybrid method spectral CT test bench and water/iodine phantom.} 
    \vspace{-2mm}
    \label{fig:TestBench}
\end{figure}

For each possible design, we established a physical model, and computed the separability index for a 0.5~mm voxel impulse of iodine and water at the center of an 80~mm cylinder of water. The final design was chosen as the one which maximized the water/iodine separability metric. The process above was also repeated for the two spectral modulation technologies acting individually. That is, the design space was constrained to cases with no filtration for kVp control design. For the k-edge filtration design, the space was constrained to designs with static kVp settings. For the hybrid design, we used the full parameter space which includes combinations of kVp settings and filtration materials.

The three optimized designs were implemented physically on the x-ray CT test bench as shown in Figure~\ref{fig:TestBench}. We constructed a water/iodine phantom using cylindrical targets designed for CT calibration. Each cylinder is approximately 27~mm in diameter and the composition of each cylinder has been designed to match the attenuation spectra of water plus some concentration of iodine. The nominal iodine concentrations are 0.0, 2.5, 5.0, 7.5, 10.0, 15.0, and 20.0~mg/mL and they are arranged as shown in Figure~\ref{fig:TestBench}. We used a 2-dimensional fan-beam system geometry with a source-to-axis distance of 831.58~mm and a source-to-detector distance of 1766.54~mm. The detector pixel size is 0.278 $\times$ 0.278 mm and the central 60 detector rows were binned to produce the one-dimensional projection measurements for each view. 

Exposure settings were calibrated using preliminary scans for each design and approximating photon counts according to the variance in the gain measurements. A voxelized approximation of the water/iodine phantom was used to approximate the dose attenuated by the phantom as defined by \eqref{eq:dose}. The target exposure was then established in such a way that the predicted dose attenuated by the phantom was normalized to 1~mJ and the ratio of exposures was matched to the design optimization results. The source mAs was scaled in proportion to the ratio between the target exposure and initial exposure estimates. The system spectral sensitivity was calibrated by scanning a phantom containing known concentrations of water and iodine and fitting a parameterized spectral model.
%%% Good but shortened...
%, reconstructing images via filtered back projection, manually segmenting and labeling basis material densities, and forward-projecting to establish a ground-truth reference for basis material line integrals. Finally, we tuned our spectral model to match the measured transmissivity. This process closely resembles standard methods of beam-hardening correction in single energy CT systems.

After calibration was complete, we scanned the water/iodine phantom using each of the three optimized designs. Two-dimensional material density images with 200~$\times$~200 voxels of size 0.5~mm were reconstructed via MBMD using 1000 iterations of the separable quadratic surrogates algorithm described in \cite{tilley2018model}. We also ran a standard model-based iterative reconstruction to estimate attenuation. This was accomplished using the same polyenergetic model used for MBMD but with a single-material (water) basis. To evaluate image quality, we computed means and cross-material covariances in 7~ROIs centered on each cylinder for the water and iodine basis material density images resulting from MBMD. We also computed variance in the same ROIs for the MBIR reconstructed image results.

\begin{figure*}[ht]
    \centering
    \includegraphics[trim={55mm 50mm 60mm 10mm},clip,width=0.58\textwidth]{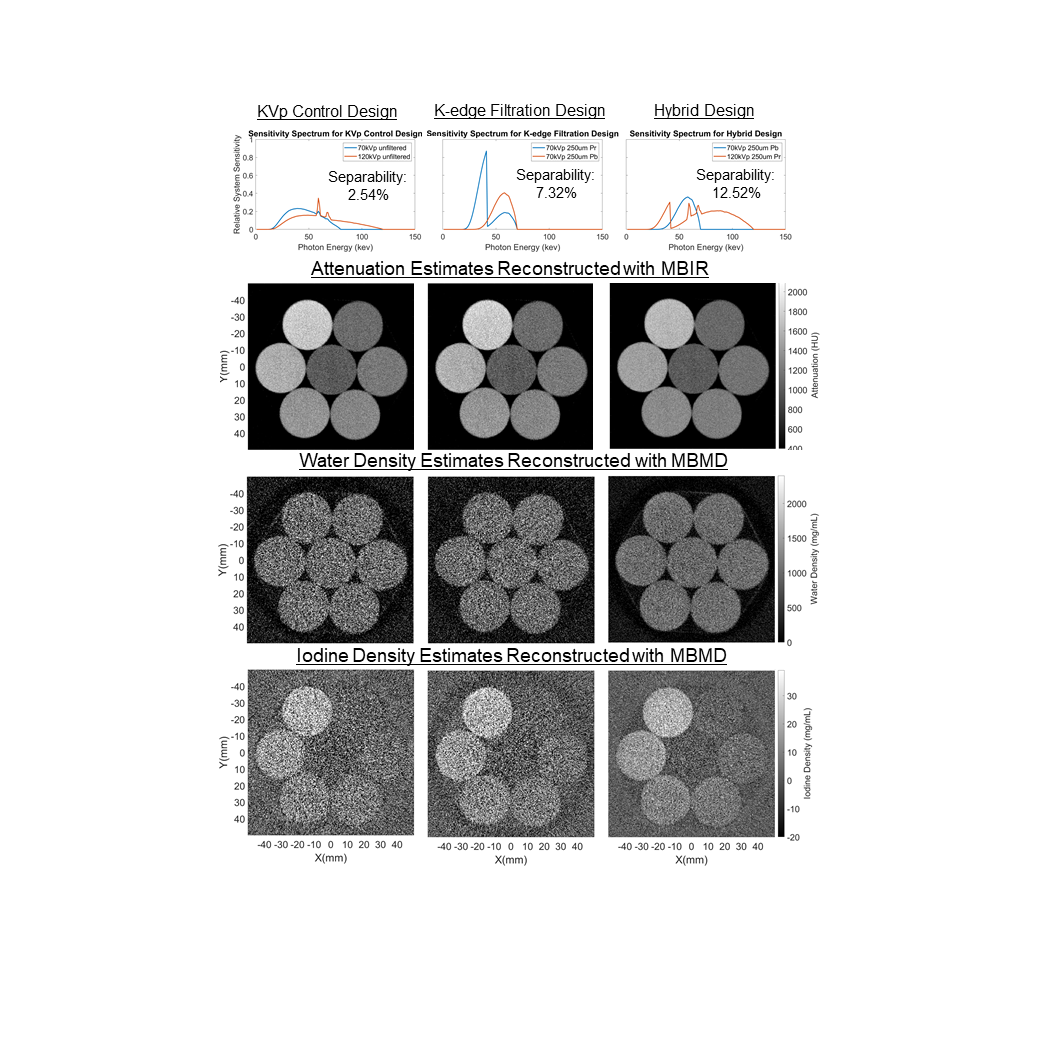}
    \includegraphics[trim={95mm 48mm 90mm 45mm},clip,width=0.35\textwidth]{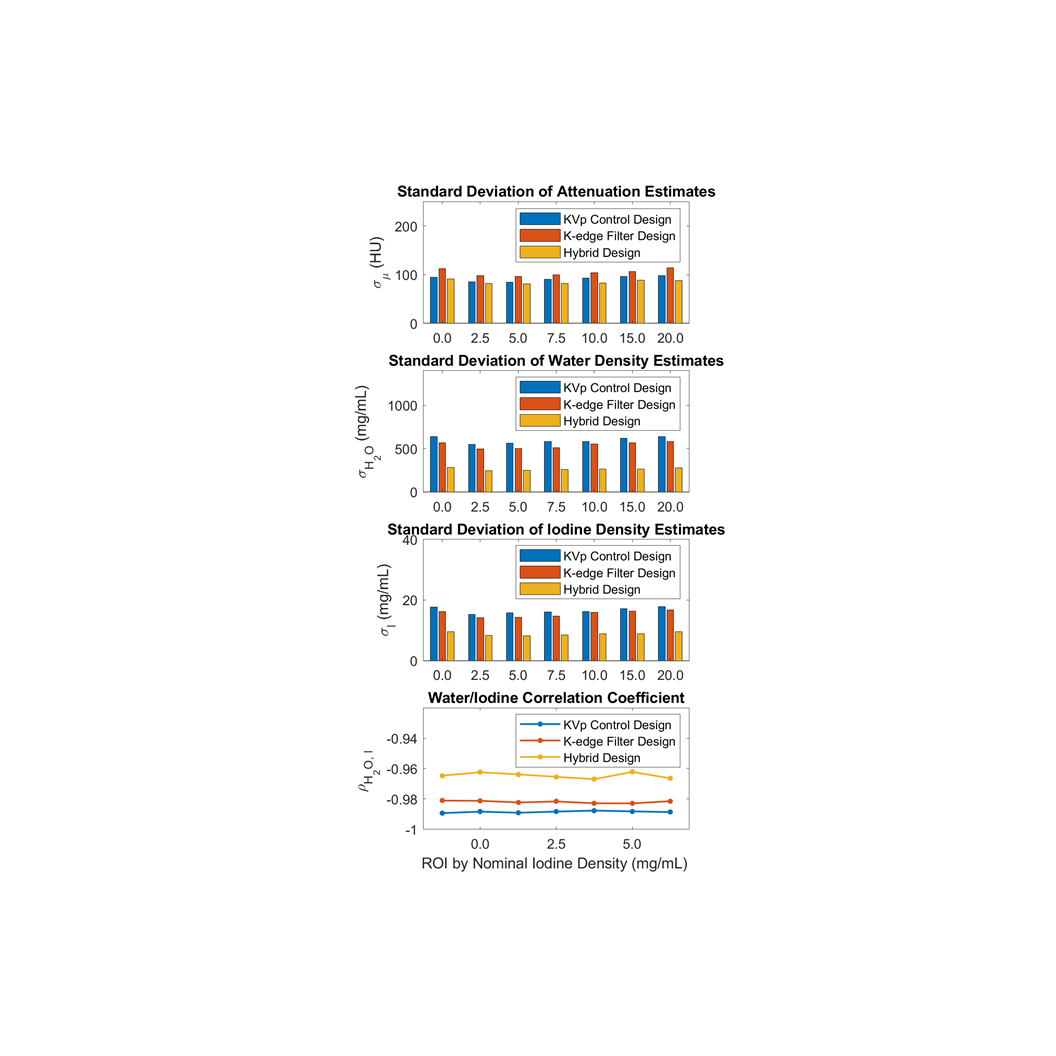}
    \caption{Spectral sensitives for each design, MBIR and MBMD reconstruction results, and covariance metrics by ROI.} 
    \vspace{-5mm}
    \label{fig:ImageResults}
\end{figure*}

\vspace{-2mm}
\section{Results}
\vspace{-1mm}

The optimized kVp control design was found to be 48.47\% of photons at 70~kVp and 51.53\% of photons at 120kVp. This design results in a water/iodine separability index of 2.54\%. This result in not particularly surprising, since this is the largest possible kVp separation, matching intuition that spectra which are very different from one another enable greater material separability. The optimized k-edge filtration design was found to be 58.16\% of photons at 70~kVp with a 250~$\mu$m praseodymium filter and 41.84\% of photons at 70~kVp with a 250~$\mu$m lead filter. This design results in a water/iodine separability index of 7.32\%. The optimized hybrid design was found to be 34.88\% of photons at 70~kVp with a 250$\mu$m lead filter and 65.12\% of photons at 120~kVp with a 250~$\mu$m praseodymium filter. This design results in a water/iodine separability index of 12.52\%.

%%% WEB - Make sure to include a reference to the Figure
The sample mean of estimated density for all ROIs was within 10\% of the nominal value for both water and iodine for all three designs. As shown in Figure \ref{fig:ImageResults} sample variance in the attenuation image estimated via MBIR is comparable for all three designs. This is expected because the total dose absorbed by the phantom was normalized for each design. The sample covariance in the water and iodine density images estimated with MBMD shows that the noise is much lower for the hybrid design than for either of the spectral modulation technologies acting individually. The correlation coefficient is also closer to zero for the hybrid design than either of the individual designs.

\vspace{-2mm}
\section{Conclusion}
\vspace{-1mm}

In this work we apply a previously developed quantitative metric of material separability based on the Fisher information matrix. We have shown that this metric can be used to optimize spectral CT system design for higher sensitivity and demonstrated efficacy in a physical system. Furthermore, the results of the imaging study show that spectral CT systems which use a combination of multiple spectral modulation technologies have the potential for improved performance with respect to designs using the constituent individual technologies. 

There are several limitations in this preliminary work which should be acknowledged. First, we used a relatively small phantom compared to the size of a human patient. Material decomposition will be a more poorly conditioned problem for larger objects. Additionally, we did not incorporate a scatter model. For low-contrast applications, scatter and other systemic biases are a concern and must be addressed. These topics are the subject of ongoing studies.

Despite these limitations, the results suggest that hybrid design is a promising strategy for spectral CT with the potential to overcome low-concentration limits of more traditional single-technology spectral methods. While this work has concentrated on two more traditional spectral technologies, other combinations including other source-side modulation schemes or  energy-sensitive detectors (such as dual-layer or photon counting detectors) could provide additional advantages. Jointly optimizing such hybrid systems for material separability may potentially have a significant benefit for clinical applications involving high sensitivity to low contrast concentrations, and will be investigated in future work.

\small

\vspace{-2mm}

\section*{Acknowledgements}

\vspace{-2mm}

This work was supported, in part, by NIH grant R21EB026849.

% \section*{References}
\vspace{-3mm}
\bibliography{report}

\begin{thebibliography}{10}
\providecommand{\url}[1]{#1}
\csname url@samestyle\endcsname
\providecommand{\newblock}{\relax}
\providecommand{\bibinfo}[2]{#2}
\providecommand{\BIBentrySTDinterwordspacing}{\spaceskip=0pt\relax}
\providecommand{\BIBentryALTinterwordstretchfactor}{4}
\providecommand{\BIBentryALTinterwordspacing}{\spaceskip=\fontdimen2\font plus
\BIBentryALTinterwordstretchfactor\fontdimen3\font minus
  \fontdimen4\font\relax}
\providecommand{\BIBforeignlanguage}[2]{{%
\expandafter\ifx\csname l@#1\endcsname\relax
\typeout{** WARNING: IEEEtran.bst: No hyphenation pattern has been}%
\typeout{** loaded for the language `#1'. Using the pattern for}%
\typeout{** the default language instead.}%
\else
\language=\csname l@#1\endcsname
\fi
#2}}
\providecommand{\BIBdecl}{\relax}
\BIBdecl

\bibitem{sauter2018dual}
Sauter, A.~P \emph{et~al.}, \emph{European journal of radiology}, vol. 104,
  2018.

\bibitem{faby2015performance}
Faby, S \emph{et~al.}, \emph{Medical physics}, vol.~42, no.~7, pp. 4349--4366,
  2015.

\bibitem{kawamoto2018assessment}
Kawamoto, S \emph{et~al.}, \emph{Abdominal Radiology}, vol.~43, no.~2, 2018.

\bibitem{xu2009dual}
Xu, D \emph{et~al.}, in \emph{Medical Imaging 2009: Physics of Medical
  Imaging}, vol. 7258.\hskip 1em plus 0.5em minus 0.4em\relax Int. Soc. for
  Optics and Photonics, 2009.

\bibitem{gang2018image}
Gang, G.~J \emph{et~al.}, \emph{Medical Physics}, vol.~45, no.~1, pp. 144--155,
  2018.

\bibitem{tivnan2019physical}
Tivnan, M \emph{et~al.}, in \emph{SPIE Medical Imaging, 2019}, 2019.

\bibitem{ma2020high}
Ma, Y.~Q \emph{et~al.}, in \emph{Conference proceedings. International
  Conference on Image Formation in X-Ray Computed Tomography}, vol. 2020, 2020.

\bibitem{taguchi2013vision}
Taguchi, K and Iwanczyk, J.~S, \emph{Medical physics}, vol.~40, no.~10, 2013.

\bibitem{primak2009improved}
Primak, A \emph{et~al.}, \emph{Medical physics}, vol.~36, no.~4, pp.
  1359--1369, 2009.

\bibitem{tivnan2020combining}
Tivnan, M \emph{et~al.}, in \emph{Medical Imaging 2020: Physics of Medical
  Imaging}, vol. 11312.\hskip 1em plus 0.5em minus 0.4em\relax International
  Society for Optics and Photonics, 2020.

\bibitem{tivnan2019optimized}
Tivnan, M and Stayman, J.~W, in \emph{Fully 3D}, vol. 11072, 2019, p. 1107211.

\bibitem{tivnan2020design}
Tivnan, M \emph{et~al.}, \emph{arXiv preprint arXiv:2010.07483}, 2020.

\bibitem{wang2019generalized}
Wang, W \emph{et~al.}, in \emph{Medical Physics}, 2019, pp. E257--E257.

\bibitem{wang2019local}
Wang, W and Webster, J, in \emph{Fully3D}.\hskip 1em plus 0.5em minus
  0.4em\relax SPIE, 2019, p. 110720Z.

\bibitem{tilley2018model}
Tilley~II, S.~W \emph{et~al.}, \emph{Physics in medicine and biology}, 2018.

\bibitem{tivnan2020preconditioned}
Tivnan, M \emph{et~al.}, 2020.

\end{thebibliography}
\bibliographystyle{spiebib-abbr} % makes bibtex use spiebib.bst

\end{document}